\documentclass[10pt]{article}
\usepackage{times}
\usepackage{graphicx}
\usepackage{latexsym}
\usepackage{amssymb}
\newtheorem{theorem}{Theorem}
\newtheorem{lemma}{Lemma}

\newcommand{\ie}{{\em i.e.,}\ }
\newcommand{\eg}{{\em e.g.,}\ }

\newenvironment{proof}{\noindent {\em Proof}:\\}{\ $\blacksquare$}

\title{Scale Invariance and Symmetry Relationships\\
In Non-Extensive Statistical Mechanics\\
}

\author{Mark Fleischer\\
Information Operations Group\\Johns Hopkins University\\
Applied Physics Laboratory\\ 11100 Johns Hopkins Road\\
Laurel, Maryland 20723-6099\\
email: Mark.Fleischer@jhuapl.edu\\
\copyright\ 2004 by Mark Fleischer}%

\date{December 10, 2004}
\begin{document}
\maketitle
\bibliographystyle{plain}
\begin{abstract}
This article extends results described in a recent article
detailing a structural scale invariance
property of the simulated annealing (SA) algorithm.  These
extensions are based on generalizations of the SA algorithm based
on Tsallis statistics 
and a non-extensive form of
entropy.  These scale invariance properties show how arbitrary
aggregations of energy levels retain certain mathematical
characteristics. In applying the non-extensive forms of
statistical mechanics to illuminate these scale invariance
properties, an interesting energy transformation is revealed that
has a number of potentially useful applications.  This energy
transformation function also reveals a number of symmetry
properties.  Further extensions of this research indicate how this
energy transformation function relates to power law distributions
and potential application for overcoming the so-called ``broken
ergodicity'' problem prevalent in many computer simulations of
critical phenomena.
\end{abstract}

\section{Introduction}\label{sec:intro}
Recent advances in statistical mechanics have helped to explain
the behavior of large ensembles of interacting systems.  Since the
time of Boltzmann, there has been a great deal of effort in
attempting to understand the behavior of such systems.  In his
day, a great deal of progress was made using a form of entropy
that seemed to explain the behavior of the so-called ``ideal gas''
model, a model premised on the very limited form of interaction
implied by elasticity assumptions.  Notwithstanding these
assumptions, it was able to predict, quite accurately, many
attributes of the behavior of gases in thermal equilibrium.

Despite its enormous success for these idealized systems, the
Boltzmann version of statistical mechanics has had limited success
in explaining the behavior of more complex systems where the
components interact in ways that are distinctly {\em
inelastic}---that is to say, the particles do not obey simple
Newtonian mechanics.  In these models, the particles themselves
may absorb energy during collisions in forms other than by changes
in their velocity.  The assumptions of the ideal gas model that
these particles have no internal structure therefore cannot be
assumed. Or, it may be that these particles exhibit attractive or
repulsive forces between them. In any event, their gross behavior
is not well predicted in systems governed by the Boltzmann-Gibbs
entropy formula: $S_{BG}=-k\sum_{i=1}^W \pi_i\ln \pi_i$ where the
$\pi_i$ represents the probability of the $i^{\mbox{\tiny th}}$
energy partition and $W$ represents the size (number of energy
states) of the system. These {\em complex systems} obey various
power laws and can exhibit phase transitions and related critical
phenomena where there are drastic shifts in their energy
configurations.

These power laws and criticality properties seem to be quite
diverse and ubiquitous in nature.  To help explain these
behaviors, Tsallis \cite{Tsallis88,Tsallis03} developed a new
entropy expression that forms the basis of a {\em non-extensive}
form of thermodynamics:
\begin{equation}\label{eqn:Sq}
S_q = \frac{k\left(1-\sum_{i=1}^Wp_i^q\right)}{q-1}
\end{equation}
where $k$ is a constant and $S_q$ is the entropy parameterized by
the entropic parameter $q$. In classical statistical mechanics,
entropy falls into a class of variables that are referred to as
{\em extensive} because they scale with the size of the system.
{\em Intensive} variables, such as temperature, do not scale with
the size of the system \footnote{Combine two vessels of gas each
with the same volume and pressure into another vessel of twice the
volume, and the pressure and temperature of the combined gas will
be the same as before. Energy and entropy, however, are examples
of {\em extensive} variables in that combining several sources of
either energy or entropy and you increase the total energy or
entropy.}.  Tsallis' form of entropy is {\em non-extensive}
because the entropy of the union of two independent systems is not
equal to the sum of the entropies of each system. That is, for
independent systems $A$ and $B$,
\begin{equation}\label{eqn:Sq(A+B)}
S_q(A+B) = S_q(A)+S_q(B) + \frac{(1-q)S_q(A)S_q(B)}{k}
\end{equation}
and Tsallis points out that $q=1$ recovers the extensivity
properties of the Boltzmann-Gibbs entropy formula and further,
that $\lim_{q\rightarrow 1}S_q=S_{BG}$, hence can be seen as a
generalization of Boltzmann-Gibbs \cite{Tsallis88,Tsallis03}.
Moreover, the stationary probabilities $p_i(t)$ of the system
being in some $i$ further illustrates the generalization of the
classic entropy form in that
\begin{equation}\label{eqn:limit-p-pi}
\lim_{q\rightarrow 1}p_i(t) = \pi_i(t)
\end{equation}
for all $t>0$ \cite[p.483]{Tsallis88}.

The entropic parameter $q$ controls important aspects of the
probability distribution of the energy levels making it is
possible to markedly change the nature of the thermodynamic system
being modeled. For example, when $q>1$ the stationary probability
distribution shifts from an exponential form to one with heavy
tails that gives rise to a power law distribution, or conversely,
one can increase the stationary probability of low energy states
when $q<1$, something quite useful in optimization (see \eg
\cite{Andricioaei96,Tsallis95}).

One of the best ways for studying the implications of this
non-extensive form of statistical mechanics is through the use of
computer simulations.  The simulated annealing (SA) algorithm
provides these basic simulation tools as it is, at heart, a
simulation of a thermodynamic system although it has been used
principally for solving optimization problems. Tsallis developed a
{\em generalized simulated annealing} (GSA) algorithm
\cite{Tsallis95} based on maximizing the entropy in
(\ref{eqn:Sq}).

Given these developments, it seems quite appropriate and useful to
examine certain scale invariance properties
\cite{FleischerSAscale02} of the classic SA algorithm in light of
the non-extensive form of entropy and the GSA.  These scale
invariance properties shed light on a number of behaviors and
illustrate some curious effects on the configuration space itself.
An energy transformation function is identified and used to
illustrate a number of symmetry properties and potential
applications. In addition, the behavior of parallel forms of SA,
also exhibit scale invariance properties and can be used to
illustrate certain analogies in the behavior of large ensembles of
interacting particles or systems, a basic aspect of non-extensive
statistical mechanical systems.

This article is organized as follows: Section \ref{sec:background}
provides some background into non-extensive thermodynamics,
simulated annealing, and SAs scale invariance properties.  Section
\ref{sec:non-ext-scale} develops the scale invariance properties
of the non-extensive form of SA.  These scale invariance
properties indicate the existence of an energy transformation
function.  Section \ref{sec:TransformedEnergy} describes certain
asymptotic properties of this energy transformation function that
suggest how it turns an exponential distribution into a power-law
distribution. Section \ref{sec:conclusion} provides some
discussion on the implications of the scale invariance properties
and the energy landscape transformation describes extensions of
this work and directions for future research involving Markov
Chain Monte Carlo simulation of complex systems.

\section{Background}\label{sec:background}
\subsection{Non-Extensive Thermodynamics and Generalized SA}
The GSA developed by Tsallis \cite{Tsallis95}, entails a different
form for the acceptance probabilities and stationary probabilities
because of the form of (\ref{eqn:Sq}), \ie the stationary
distribution is based on maximizing the value of $S_q$ rather than
$S_{BG}$. The latter leads to the well-known Boltzmann-Gibbs
distribution
\begin{equation}\label{eqn:def-pi}
\pi_i(t) = \frac{e^{-f_i/t}}{Z_{BG}}
\end{equation}
where $Z_{BG} = \sum_{j=1}^W e^{-f_j/t}$ is the normalizing
Boltzmann Partition Function.  Note that because we will often be
referring to the SA algorithm and the Metropolis Algorithm, the
value of the ``energy'' function will be denoted by $f_i$ which is
typically used in optimization problems to denote an objective
function value, but this can also denote some energy value.

Maximizing the non-extensive entropy subject to certain
constraints, described below, gives rise to a stationary
probability distribution
\begin{equation}\label{eqn:defTallis-pi}
p_i(t) =
\frac{\left[1+\left(\frac{q-1}{t}\right)f_i\right]^{\frac{1}{1-q}}}{Z_q}
\end{equation}
where
$Z_q=\sum_{j=1}^W\left[1+\left(\frac{q-1}{t}\right)f_j\right]^{\frac{1}{1-q}}.$
Note that $p_i(t)$ will be used throughout to represent the
stationary probability in the non-extensive case and $\pi_i(t)$
will represent the stationary probability in the classic,
extensive case.

The definition of the stationary probability $p_i(t)$ in
(\ref{eqn:defTallis-pi}) is based on the particular set of
constraints used in defining a thermodynamic system.  Three
distinct and noteworthy sets of constraints have been studied all
of which employ the standard normalizing constraint
$\sum_ip_i(t)=1$.  What distinguishes these constraint sets is the
relationships they define between probabilities and energy or
objective function values.  Tsallis \cite{Tsallis98} reports that
different forms of these constraints have a number of implications
for the non-extensive form of the probability $p_i(t)$. Here, the
form defined by Tsallis' Type 2 constraint \cite{Tsallis96} for
GSA is used where,
\begin{equation}\label{eqn:pqconstraint}
\sum_{i=1}^Wp_i^qf_i = U^{(2)}, \mbox{ a constant},
\end{equation}
whereas in his original paper, the constraint
$\sum_i\tilde{p}_if_i = U$ was used.  This later constraint was
referred to as the Type 1 constraint in \cite[p. 537]{Tsallis98}
and leads to a different stationary probability denoted here by
$\tilde{p}_i$.  It bears emphasis that the limit in
(\ref{eqn:limit-p-pi}) holds for all the different stationary
probabilities that arise from the use of the different constraints
\cite{Tsallis98}.  Initially, the notions of scale invariant
structures will be based on the Type~2 constraint and then later,
in a more useful context, the scale invariance of the Type~1
constraint will be examined.

\subsection{Classical SA Scale Invariance}
Among the many interesting aspects of the SA and Metropolis
algorithms is a scale invariance property associated with the
stationary probabilities of various states.  This scale invariance
property is manifest in the identical mathematical forms of
certain quantities involving individual states, aggregated states
of the solution space, and the aggregation of states associated
with multiple processors in an expanded state space. See
\cite{FleischerSAscale02,Fleischer_dis} for a complete
description.

One aspect of this scale invariance involves the rate change of
the stationary probability of a state $i$ with respect to
temperature $t$:
\begin{equation}\label{eqn:deriv-pi-ClassicSA}
\frac{\partial \pi_i(t)}{\partial t} =
\frac{\pi_i(t)}{t^2}\left[f_i - \langle f\rangle(t)\right]
\end{equation}
where $\langle f \rangle(t)$ is the expected objective function
value at temperature $t$.  For aggregated states $A = \{i_1,
i_2,\ldots, i_m\}$, define
\begin{equation}\label{eqn:def-piAandfA}
\pi_A(t) \equiv \sum_{i\in A}\pi_i(t),\ \ f_A(t) \equiv
\frac{\sum_{i\in A}\pi_i(t)f_i}{\pi_A(t)},
\end{equation}
the conditional expectation of the objective function value given
that the current state is in set $A$ (see
\cite[p.224]{FleischerSAscale02} for a more complete treatment).
It then follows that
\begin{equation}\label{eqn:derivpiA}
\frac{\partial \pi_A(t)}{\partial t} =
\frac{\pi_A(t)}{t^2}\left[f_A(t) - \langle f\rangle(t)\right]
\end{equation}
where the similarity of (\ref{eqn:deriv-pi-ClassicSA}) and
(\ref{eqn:derivpiA}) indicates a form of scale invariance.

Scale invariance also extends to second moments. For standard SA,
\begin{equation}\label{eqn:deriv-fbar}
\frac{\partial \langle f
\rangle(t)}{\partial t} = \frac{\partial f_\Omega(t)}{\partial t}
= \frac{\sigma^2_\Omega(t)}{t^2}
\end{equation}
where $\sigma^2_\Omega(t)$ represents the variance over the entire
solution space at temperature $t$. Scale invariance is exhibited
by the fact that
\begin{equation}\label{eqn:deriv-fAbar}
\frac{\partial f_A(t)}{\partial t} = \frac{\sigma^2_A(t)}{t^2}
\end{equation}
and $\sigma_A^2(t)$ is the variance of objective function values
at temperature $t$ conditioned on the current state being in set
$A$. See \cite[p.232-33]{FleischerSAscale02} for details.

\section{SA Scale Invariance with Non-Extensive
Entropy}\label{sec:non-ext-scale}
\subsection{The Basis of Scale Invariance}
To demonstrate scale invariance based on aggregated states in the
non-extensive case, a basis for making comparisons must be
established. To that end, the following temperature derivative of
$p_i(t)$ is calculated for the non-extensive SA case. For
notational convenience, simplicity and to ensure positivity, let
$N_i(t)\equiv
\left[1+\left(\frac{q-1}{t}\right)f_i\right]^{\frac{1}{1-q}}$
(hereinafter we will drop the $(t)$ from $N_i(t)$ to further
simplify the expressions) and taking the derivative of
(\ref{eqn:defTallis-pi}),
\begin{equation}\label{eqn:derivative-1-pi}
\frac{\partial p_i(t)}{\partial t} =
\frac{Z_q\frac{\partial}{\partial t}N_i -
N_i\frac{\partial}{\partial t}Z_q}{(Z_q)^2}.
\end{equation}
Note that
\begin{equation}
\frac{\partial  N_i}{\partial t} = \frac{\partial}{\partial
t}\left[1 + \left(\frac{q-1}{t} \right)f_i
\right]^{\frac{1}{1-q}}= \frac{f_iN_i^q}{t^2}, \label{eqn:derivNi}
\end{equation}
$Z_q = \sum_jN_j,$ and $\  p_i(t)=N_i/Z_q$ and hence
\begin{displaymath}
 \frac{\partial}{\partial t}Z_q =
 \frac{\partial}{\partial t}\sum_jN_j = \sum_j\frac{\partial
N_j}{\partial t} =\sum_j\frac{f_jN_j^q}{t^2}.
\end{displaymath}
Substituting this and (\ref{eqn:derivNi}) into
(\ref{eqn:derivative-1-pi}) yields
\begin{eqnarray}
\frac{\partial p_i(t)}{\partial t} &=&
\frac{\left(\sum_jN_j\right)\frac{f_iN^q_i}{t^2} -
\frac{N_i}{t^2}\sum_jf_jN_j^q}{(\sum_jN_j)^2}\nonumber\\
&=& \frac{p_i(t)f_iN_i^{q-1}}{t^2}-
\frac{p_i(t)\sum_jf_jN_j^q}{t^2\sum_jN_j}\nonumber\\
&=&\frac{p_i(t)}{t^2}\left[f_iN_i^{q-1}-\frac{\sum_jf_jN_j^q}{\sum_jN_j}\right]\label{eqn:deriv-p-b4fhatdef}
\end{eqnarray}
To further simplify and clarify, define
\begin{equation}\label{eqn:def-fhat}
\hat{f}_i(f_i,q,t)\equiv
f_iN_i^{q-1}=\frac{f_i}{1+\left(\frac{q-1}{t}\right)f_i}
\end{equation}
and will often be denoted simply by $\hat{f}_i$ or $\hat{f}_i(t)$
where it is understood to involve $f_i, q$ and $t$.  Substituting
(\ref{eqn:def-fhat}) into (\ref{eqn:deriv-p-b4fhatdef}) and
further simplifying yields

\begin{equation}\label{eqn:deriv-p-end}
\frac{\partial p_i(t)}{\partial t}
=\frac{p_i(t)}{t^2}\left[\hat{f}_i(t)- \langle
\hat{f}\rangle(t)\right]
\end{equation}
and which interestingly enough is analogous to
(\ref{eqn:deriv-pi-ClassicSA}). Notice that for $q=1,\
\hat{f}_i(t)=f_i,\ \forall \ t>0$.  Tsallis
\cite{Tsallis88,Tsallis03} established that $\lim_{q\rightarrow
1}p_i(t) = \pi_i(t)$, hence, in the limit as $q\rightarrow 1$,
(\ref{eqn:deriv-p-end}) $\rightarrow$
(\ref{eqn:deriv-pi-ClassicSA}). The symbol $\hat{f_i}(t)$ will be
referred to as the $q$-transformation of $f_i$.  Later on, some
interesting properties of this transformation will be examined.

\subsection{Scale Invariance from the Aggregation of States}
In similar fashion as in \cite{FleischerSAscale02}, define $p_A(t)
= \sum_{i\in A}p_i(t)$ for $A\subset \Omega$, where $\Omega$ is
the set of microstates and where $W=|\Omega|$ and
\begin{equation}\label{eqn:def-fhat-A}
\hat{f}_A(t) =\frac{\sum_{i\in A}p_i(t)\hat{f}_i(t)}{p_A(t)},
\end{equation}
the conditional expectation of $\hat{f}_i(t)$ conditioned on the
current state being in set $A$.  Taking the derivative,
\begin{equation}\label{eqn:deriv-lumped-1}
\frac{\partial p_A(t)}{\partial t} = \frac{\partial }{\partial
t}\sum_{i\in A}p_i(t) = \sum_{i\in A}\frac{\partial
p_i(t)}{\partial t}.
\end{equation}

\noindent Substituting (\ref{eqn:deriv-p-end}) and
(\ref{eqn:def-fhat-A}) into (\ref{eqn:deriv-lumped-1}) yields
\begin{eqnarray}
\frac{\partial p_A(t)}{\partial t} &=& \sum_{i\in
A}\frac{p_i(t)}{t^2}\left[\hat{f}_i(t)- \langle
\hat{f}\rangle(t)\right]\nonumber\\
&=&\sum_{i\in A}\frac{p_i(t)\hat{f}_i(t)}{t^2} -\sum_{i\in
A}\frac{p_i(t)\langle\hat{f}\rangle(t)}{t^2}\nonumber\\
&=&\frac{p_A(t)}{t^2}\sum_{i\in
A}\frac{p_i(t)\hat{f}_i(t)}{p_A(t)}-
\frac{p_A(t)\langle\hat{f}\rangle(t)}{t^2}\nonumber\\
&=&\frac{p_A(t)}{t^2}\left[\hat{f}_A(t) -
\langle\hat{f}\rangle(t)\right]\label{eqn:deriv-pA}
\end{eqnarray}
where for aggregated states, (\ref{eqn:deriv-pA}) has a similar
mathematical structure as (\ref{eqn:deriv-p-end}), hence exhibits
a scale invariance property the foundation of which is based on
the energy transformation function $\hat{f}_i(f_i,q,t)$.

\subsubsection{Scale Invariance in Second Moments}

Scale invariance in the non-extensive form of SA for second
moments is indicated in the following where again the
$q$-transformation of $f_i$ and $f_A(t)$ is used. First, the
parallels to the classic case is illustrated.  Thus,
\begin{eqnarray}
\frac{\partial \langle \hat{f} \rangle(t)}{\partial t}& =&
\frac{\partial \hat{f}_\Omega (t)}{\partial t}=
\frac{\partial}{\partial t} \left[
\sum_{i\in\Omega}p_i(t)\hat{f}_i(t)\right]\nonumber\\
&=&\sum_{i\in\Omega}\frac{\partial}{\partial
t}\left[p_i(t)\hat{f}_i(t)\right]\nonumber\\
&=&\sum_{i\in\Omega}\left[\frac{\partial p_i(t)}{\partial
t}\hat{f}_i(t) + p_i(t)\frac{\partial \hat{f}_i(t)}{\partial
t}\right].\label{eqn:deriv-<fhat>Omega}
\end{eqnarray}

\noindent Substituting (\ref{eqn:deriv-p-end}) into the first part
of (\ref{eqn:deriv-<fhat>Omega}) and simplifying yields
\begin{equation}
\frac{\partial \langle \hat{f} \rangle(t)}{\partial t}=
\sum_{i\in\Omega}\frac{p_i(t)\hat{f}^2_i(t)}{t^2}-
\sum_{i\in\Omega}\frac{\hat{f}_\Omega(t)p_i(t)\hat{f}_i(t)}{t^2}+
\sum_{i\in\Omega}p_i(t)\frac{\partial\hat{f}_i(t)}{\partial
t}\label{eqn:threetermsofderiv-<fhat>Omega}.
\end{equation}
Noting the form of the first two terms in
(\ref{eqn:threetermsofderiv-<fhat>Omega}) and the fact that in the
third term
\begin{equation}\label{eqn:derivhatf}
\frac{\partial \hat{f}_i(t)}{\partial t} =
\left(\hat{f}_i(t)\right)^2\left(\frac{q-1}{t^2}\right)
\end{equation}
and substituting into (\ref{eqn:threetermsofderiv-<fhat>Omega})
and dropping the $(t)$ for notational clarity and adding the
symbol $\Omega$ to denote expectations over the entire state space
yields
\begin{eqnarray}
\frac{\partial {\langle \hat{f} \rangle}_\Omega}{\partial t} &=&
\frac{{\langle
\hat{f}^2\rangle}_\Omega-{\langle\hat{f}\rangle}_\Omega^2+(q-1){\langle
\hat{f}^2\rangle}_\Omega}{t^2}\nonumber\\
&=&\frac{\hat{\sigma}_\Omega^2}{t^2} +
\frac{(q-1){\langle\hat{f}^2\rangle}_\Omega}{t^2}\label{eqn:deriv-fhat-Omega}
\end{eqnarray}
where $\hat{\sigma}_\Omega^2$ represents the variance of the
$q$-transformed objective function values over the entire
objective function space (at temperature $t$).  Obviously, this
expression is slightly different from the classic SA case as
indicated in (\ref{eqn:deriv-fbar}).  Note that
$\lim_{q\rightarrow 1}\hat{\sigma}_\Omega^2 = \sigma_\Omega^2$ so
that the expressions in (\ref{eqn:deriv-fhat-Omega}) and
(\ref{eqn:deriv-fbar}) become equivalent.

The expression in (\ref{eqn:deriv-fhat-Omega}) quite clearly shows
the effects of the value of $q$---values greater (less) than 1
increase (decrease) the rate of change of the expected objective
function values ($q$-transformed values).  Eq.
(\ref{eqn:deriv-fhat-Omega}) also provides the basis for another
form of scale invariance.  Thus, after going through similar steps
as in (\ref{eqn:deriv-<fhat>Omega}) through
(\ref{eqn:deriv-fhat-Omega}) we get
\begin{equation}\label{eqn:derivfhatAraw}
\frac{\partial \hat{f}_A}{\partial t}= \frac{\sum_{i\in A}p_i
\hat{f}_i^2}{t^2p_A} - \frac{\hat{f}_A^2}{t^2}+
\left(\frac{q-1}{t^2} \right)\frac{\sum_{i\in
A}p_i\hat{f}^2_i}{p_A}.
\end{equation}
Noting that the first and third terms indicate conditional
expectations conditioned on the current state being in set $A$,
then (\ref{eqn:derivfhatAraw}) can be re-written in the convenient
notation
\begin{eqnarray}
\frac{\partial \hat{f}_A}{\partial t}&=& \frac{{\langle
\hat{f}^2\rangle}_A - {\langle\hat{f}\rangle}_A^2 +(q-1){\langle
\hat{f}^2\rangle}_A}{t^2}\nonumber\\
&=&\frac{\hat{\sigma}^2_A}{t^2}+ \frac{(q-1){\langle
\hat{f}^2\rangle}_A}{t^2}\label{eqn:derivfhatAend}
\end{eqnarray}
where (\ref{eqn:derivfhatAend}) is clearly analogous to
(\ref{eqn:deriv-fhat-Omega}) and so exhibits a form of scale
invariance.  Again, in the limit as $q\rightarrow 1$ both of these
equations correspond to the scale invariance of standard
(Boltzmann-Gibbs) SA as indicated in (\ref{eqn:deriv-fbar}) and
(\ref{eqn:deriv-fAbar}). What is interesting however is the fact
that the terms involving $(q-1)$ also scale with the aggregated
set $A$ in the non-extensive case.

\subsection{Scale Invariance from the Aggregation of Processors}
\subsubsection{The Classic SA Case}
Fleischer \cite{Fleischer_JOH,Fleischer_MIC97,Fleischer_MIC95}
describes another form of scale invariance based on the
aggregation parallel and independent processes each running the SA
algorithm independently of one another.  This type of aggregation
contrasts sharply with the aggregation of states and hence
directly affects the nature of the scale invariance and how the
algorithm itself functions. In this respect, one may view the SA
algorithm as a single processor or {\em particle} that
probabilistically visits a single state at each iteration of the
algorithm.  Thus, when a set of states is aggregated and fixed,
the particle visits the aggregated state probabilistically and in
a way that dictates how the stationary probability and objective
function value should reasonably be defined, \ie based on the sum
of the stationary probabilities and the conditional expectation of
the objective function respectively as given in
(\ref{eqn:def-piAandfA}).

In aggregating $p$ independent processors however, the state space
$\Omega$ becomes $\Omega^p$ and the stationary probability and
objective function values associated with the set of processors
must be defined differently and requires the following
definitions. First, define a state in the product space spanned by
$p$ processors as $i_1i_2\ldots i_p$.  In classic SA scale
invariance
\begin{equation}\label{eqn:def-pi-prodspace}
\pi_{i_1i_2\ldots i_p}(t)=\prod_{m=1}^p\pi_{i_m}(t)
\end{equation}
and is based on the concept of the joint probability of
independent processors.  Similarly, the objective function
associated with a set of $p$ independent particles is simply the
sum of each particle's objective function value as opposed to the
conditional expectation, and defined by
\begin{displaymath}
f_{i_1i_2\ldots i_p}\equiv \sum_{m=1}^p f_{i_m}.
\end{displaymath}
Scale invariance is indicated by
\begin{displaymath}
\pi_{i_1i_2\ldots i_p}(t)= \frac{e^{-f_{i_1i_2\ldots
i_p}/t}}{\sum_{j_p}\cdots\sum_{j_1} e^{-f_{j_1j_2\ldots j_p}/t}}
\end{displaymath}
(see \cite{FleischerSAscale02,Fleischer_MIC97}) which has the same
form as (\ref{eqn:def-pi}) and the fact that the temperature
derivative is proportional to the difference between a function of
the objective function value and its expectation:
\begin{equation}\label{eqn:deriv-pi-prod}
\frac{\partial \pi_{i_1i_2\ldots i_p}(t)}{\partial t}=
\frac{\pi_{i_1i_2\ldots i_p}(t)}{t^2}\left[f_{i_1i_2\ldots i_p} -
{\langle f \rangle}_{\Omega^p}(t) \right]
\end{equation}
\cite[p.222]{FleischerSAscale02} where the expectation is over all
the states of the product space.

\subsubsection{The Non-Extensive SA Case}
Taking cues from this earlier work, define the stationary
probability of state $i_1i_2\ldots i_p$ spanned by $p$ independent
processors and based on the Tsallis entropy by
\begin{displaymath}
p_{i_1i_2\ldots i_p}(t) \equiv \prod_{m=1}^pp_{i_m}(t).
\end{displaymath}
Simplifying using the case of $p=2$ (the following expressions are
readily extended to the more general case) and letting $a=(q-1)/t$
for notational clarity, the joint probability of a state $i_1i_2$
is
\begin{eqnarray}
p_{i_1i_2}(t)&=&\frac{[1+af_{i_1}]^{\frac{1}{1-q}}[1+af_{i_2}]^{\frac{1}{1-q}}}{Z_q^2}\nonumber\\
&=&\frac{[1+a(f_{i_1}+f_{i_2}+af_{i_1}f_{i_2})]^{\frac{1}{1-q}}}{Z_q^2}\nonumber\\
&=&\frac{[1+a\tilde{f}_{i_1i_2}(t)]^{\frac{1}{1-q}}}{Z_q^2}\label{eqn:pxy}
\end{eqnarray}
where $\tilde{f}_{i_1i_2}(t) \equiv f_{i_1}+f_{i_2} +
af_{i_1}f_{i_2}$ and where it is also the case that
\begin{eqnarray*}
Z_q^2& =& \left(\sum_{i=1}^W\left[1+ af_i\right]^{\frac{1}{1-q}}\right)^2\\
&= &\sum_{i_1=1}^W\sum_{i_2=1}^W\left[1 + a\tilde{f}_{i_1i_2}(t)
\right]^{\frac{1}{1-q}}.
\end{eqnarray*}
Thus, the form of $p_{i_1i_2}(t)$ in the product space closely
follows that of $p_i(t)$ with the advantage that the
non-extensivity property is exemplified by $\tilde{f}_{i_1i_2}(t)$
which is somewhat similar to the entropy relationships in
(\ref{eqn:Sq(A+B)}).  Note that for $q=1, \ \tilde{f}_{i_1i_2}(t)
= f_{i_1i_2} = f_{i_1} + f_{i_2}$ for all $t$.  Thus, in the limit
as $q\rightarrow 1$, (\ref{eqn:pxy}) is equivalent to the
Boltzmann-Gibbs case in (\ref{eqn:def-pi-prodspace}) (recall
(\ref{eqn:limit-p-pi})).

Examining the derivative of the stationary probability using
(\ref{eqn:deriv-p-end}) also reflects a scale invariance property
and yields
\begin{eqnarray*}
\frac{\partial p_{i_1i_2}(t)}{\partial
t}&=&\frac{\partial}{\partial t}\left[p_{i_1}\kern
-2pt(t)p_{i_2}\kern -2pt(t)\right]\nonumber\\
&=&p_{i_1}\kern -2pt(t)\frac{p_{i_2}\kern
-2pt(t)}{t^2}\left[\hat{f}_{i_2} -
{\langle\hat{f}\rangle}_\Omega \right] \nonumber\\
& & \hspace{10pt} + p_{i_2}\kern -2pt(t)\frac{p_{i_1}\kern
-2pt(t)}{t^2}\left[\hat{f}_{i_1} - {\langle\hat{f}\rangle}_\Omega
\right]\nonumber\\
&=& \frac{p_{i_1i_2}\kern -2pt(t)}{t^2}\left[\hat{f}_{i_1i_2} -
{\langle\hat{f}\rangle}_{\Omega^2} \right]
\end{eqnarray*}
and in general it follows that
\begin{equation}\label{eqn:deriv-p-prod}
\frac{\partial p_{i_1\cdots i_p}\kern -2pt(t)}{\partial t} =
\frac{p_{i_1\cdots i_p}\kern -2pt(t)}{t^2}\left[\hat{f}_{i_1\cdots
i_p} - {\langle\hat{f}\rangle}_{\Omega^p} \right]
\end{equation}
where $\hat{f}_{i_1\cdots i_p} = \sum_{m=1}^p\hat{f}_{i_m}$ and
${\langle\hat{f}\rangle}_{\Omega^p} =
p{\langle\hat{f}\rangle}_\Omega$ and equals the expectation of
$\hat{f}_{i_1\cdots i_p}$.  Note that (\ref{eqn:deriv-p-prod})
again has a factor equal to the difference between a function of
the objective function values and its expectation and has the same
structure as in (\ref{eqn:deriv-pi-ClassicSA}),
(\ref{eqn:derivpiA}), (\ref{eqn:deriv-p-end}),
(\ref{eqn:deriv-pA}) and (\ref{eqn:deriv-pi-prod}).

Finally, let us examine the temperature derivative of ${\langle
\hat{f}\rangle}_{\Omega^p}$ and see how it compares to
(\ref{eqn:deriv-fhat-Omega}) and (\ref{eqn:derivfhatAend}).  Thus,
\begin{displaymath}
\frac{\partial {\langle \hat{f}\rangle}_{\Omega^p}}{\partial t}=
\frac{\partial}{\partial t}\sum_{i_1\cdots i_p}p_{i_1\cdots
i_p}\kern -2pt(t)\hat{f}_{i_1\cdots i_p}\kern -2pt(t).
\end{displaymath}
Using the results from (\ref{eqn:deriv-<fhat>Omega}) through
(\ref{eqn:derivhatf}) and noting that $\hat{f}_{i_1\cdots i_p}(t)$
is a sum of terms,
\begin{displaymath}
\frac{\partial \hat{f}_{i_1\cdots i_p}}{\partial t}=
\left(\frac{q-1}{t^2}\right)\sum_{m=1}^p\hat{f}_{i_m}^{\kern 1.5pt
2} = \left(\frac{q-1}{t^2}\right) \hat{f}_{i_1\cdots i_p}^{\kern
1.5pt 2}
\end{displaymath}
yields
\begin{eqnarray}
\frac{\partial {\langle \hat{f} \rangle}_{\Omega^p}}{\partial t}
&=& \frac{{\langle
\hat{f}^2\rangle}_{\Omega^p}-{\langle\hat{f}\rangle}^2_{\Omega^p}+(q-1){\langle
\hat{f}^2\rangle}_{\Omega^p}}{t^2}\nonumber\\
&=&\frac{\hat{\sigma}_{\Omega^p}^2}{t^2} +
\frac{(q-1){\langle\hat{f}^2\rangle}_{\Omega^p}}{t^2}\label{eqn:deriv-fhat-Omega^p}
\end{eqnarray}
where $\hat{\sigma}^2_{\Omega^p}$ represents the variance of the
sums $\hat{f}_{i_1\cdots i_p}(t)$ (again, bear in mind that the
dependence on $t$ is not indicated). Eq.
(\ref{eqn:deriv-fhat-Omega^p}) has the same form as
(\ref{eqn:deriv-<fhat>Omega}) and (\ref{eqn:derivfhatAend})
indicating a form of scale invariance. What is remarkable however
is that in aggregating processors, the objective function is
defined as the {\em sum} of the objective function values of each
processor whereas in aggregating states, the objective function is
defined as the {\em conditional expectation} of objective
functions, yet they have the same form and in the limit as
$q\rightarrow 1$ are equivalent to the aggregation of processors
in classic SA.

\subsection{Consistency} Fleischer \cite{FleischerSAscale02}
describes the concept of consistency in the scale invariant
structure of SA based on the definition of the aggregated
objective function $f_A$.  This consistency property states that
one can use identical means to define the objective function (here
we describe it in terms of an energy level) associated with the
aggregation of two or more {\em aggregated\/} states $A$ and $B$.
Thus,
\begin{eqnarray*}
f_{A\cup B}(t) &= &\sum_{i \in A\cup B} \frac{\pi_i(t)f_i
}{\pi_{A\cup B}(t)}\nonumber\\
&=&\sum_{i \in A} \frac{\pi_i(t)f_i }{\pi_{A\cup B}(t)}+ \sum_{i
\in B} \frac{\pi_i(t)f_i}{\pi_{A\cup B}(t)}\nonumber\\
&=&\frac{\pi_A(t)f_A(t) + \pi_B(t)f_B(t)}{\pi_A(t) +
\pi_B(t)}
\end{eqnarray*}
The definition of the objective function for state $A\cup B$ has
the same form as the definition of aggregated states $A$ and $B$
and thus the scaling phenomenon in these definitions is, indeed,
invariant on all levels of scale.  It is easy to see that this
same consistency property holds for the transformed objective
function (energy level) $\hat{f}_{A\cup B}$ based on values
$\hat{f}_A$ and $\hat{f}_B$ in the non-extensive case. In
non-extensive systems however, the concept of consistency extends
a bit further than in the classical case because of the energy
transformation function.

The alert reader may have wondered, for example, if the definition
of $\hat{f}_A(t)$ for some aggregation of energy levels $A$
defined in (\ref{eqn:def-fhat-A}) could be equated to the
$q$-transformation of some value $f_A(t)$ obtained by the weighted
average of the {\em un}transformed values $f_i$. In other words,
\begin{equation}\label{eqn:fhatconsistency}
\hat{f}_A \ =\ \frac{\sum_{i\in A}p_i\hat{f}_i}{p_A}\
\stackrel{?}{=}\ \frac{f_A}{1 + af_A}.
\end{equation}
where
\begin{equation}\label{eqn:def-fA}
f_A = \frac{\sum_{i\in A}p_if_i}{p_A}.
\end{equation}
In general, the weighted sum of transformed energy values is not
equal to the transformation of the weighted sum of energy values.
The exception is when $q=1$ or the temperature is infinite for
then $a=0$ and $\hat{f}_A = f_A$ and $p_A = \pi_A$ and the notion
of consistency described in \cite{FleischerSAscale02} holds
because the relationships here are in effect the same as in
\cite{FleischerSAscale02}.  The following theorem shows that these
quantities are approximately equal for any given {\em finite}
temperature for values of $f_i>0$ or $a > 0$ sufficiently high.
\begin{theorem}\label{thm:consistency}
Let $\hat{f}_A$ and $f_A$ be defined as in
(\ref{eqn:fhatconsistency}) and (\ref{eqn:def-fA}), respectively
with $a>0$. Then for values of $f_{i\in A}$ or $a$ sufficiently
high,
\begin{displaymath}
\hat{f}_A \ \approx\ \frac{f_A}{1+af_A}.
\end{displaymath}
\end{theorem}

\begin{proof}
This statement is proved by showing that for any $\epsilon > 0$,
there is an energy value $f^*=\displaystyle\min_{i\in A}\{f_i \}$
or an $a > 0$ sufficiently high such that
\begin{equation}\label{eqn:difference_in_epsilon}
\left|\frac{\sum_{i\in A}p_i\hat{f}_i}{p_A} - \frac{f_A}{1+\,af_A}
\right| < \epsilon.
\end{equation}
For notational clarity, let $\omega_i = \frac{p_i}{p_A}$ and hence
$\sum_{i\in A}\omega_i=1$.  Rewriting
(\ref{eqn:difference_in_epsilon}) and taking the limit as
$f^*\rightarrow\infty$ we obtain
\begin{eqnarray*}
\lefteqn{\lim_{f^*\rightarrow\infty}\left|\sum_{i\in
A}\frac{\omega_if_i}{1+af_i} - \frac{\sum_{i\in
A}\omega_if_i}{1+\,a\sum_{i\in A}\omega_if_i} \right|}\hspace{40pt}\\
&= & \left|\lim_{f^*\rightarrow\infty}\left(\sum_{i\in
A}\frac{\omega_if_i}{1+af_i} - \frac{\sum_{i\in
A}\omega_if_i}{1+\,a\sum_{i\in A}\omega_if_i}\right) \right|\\
&= &\left|\left(\lim_{f^*\rightarrow\infty}\sum_{i\in
A}\frac{\omega_i}{\frac{1}{f_i}+a}\right)-
 \left(
\lim_{f^*\rightarrow\infty}\frac{1}{\left(\frac{1}{\sum_{i\in
A}\omega_if_i}\right)+\,a}\right) \right |\\
&=&\left|\frac{1}{a} - \frac{1}{a}\right | = 0
\end{eqnarray*}
since for all $i\in A, f_i \geq f^*$.  For $a > 0$ sufficiently
high, both terms have limits of zero.  Consequently, for all
$\epsilon > 0$, there exists an $f^* $ or $a > 0$ such that
(\ref{eqn:difference_in_epsilon}) holds.
\end{proof}

Theorem \ref{thm:consistency} shows that in the high energy
regions of the energy landscape it makes no significant difference
if energy levels are aggregated first to obtain the value of $f_A$
and then transforming that value to obtain $\hat{f}_A$ or we
transform the values $f_i$ first to obtain the $\hat{f}_i$ and
{\em then} aggregate them to obtain $\hat{f}_A$---either way, they
both yield values close to $1/a$.

\section{Symmetry Relationships}\label{eqn:symmetry}
Tsallis observes a symmetry relationship in using the different
constraints described earlier in Section \ref{sec:background} and
in text surrounding (\ref{eqn:pqconstraint}) \cite{Tsallis98}.
Tsallis' original incarnation of non-extensivity was based on his
Type 1 constraint $\sum_i\tilde{p}_if_i = U^{(1)}$ (see
\cite{Tsallis98}) for stationary probability
\begin{equation}\label{eqn:def-ptilde}
\tilde{p}_i(t) = \frac{\left[1 + \left(\frac{1-q}{t}\right)f_i
\right]^{\frac{1}{q-1}}}{Z_q^\prime}
\end{equation}
where $Z_q^\prime$ is the obvious normalizing constant (in physics
nomenclature, the $U^{(1)}$ refers to the expected value of the
eigenvalues of the Hamiltonian of the system although here, we can
simply refer to it as the expected value of some energy).  Tsallis
points out that the form of $\tilde{p}_i$ is essentially the same
as that of $p_i(t)$ except that $1-q$ replaces every occurrence of
$q-1$ and vice versa {\em including in the exponents} (compare
(\ref{eqn:def-ptilde}) with (\ref{eqn:defTallis-pi})).  This
interesting fact provides hints of additional symmetry
relationships. In this section, these types of symmetry
relationships are further explored in light of the scale
invariance properties already developed.

\subsection{Scale Invariance Using Other Constraints}
It is quite reasonable to ask whether the constraint
$\sum_i\tilde{p}_if_i=U^{(1)}$ also leads to scale invariant forms
and other symmetry relationships such as those described above.
Define the energy transformation
\begin{equation}\label{eqn:def-ftilde}
\tilde{f}_i(f_i,q,t) \equiv
\frac{f_i}{1+\left(\frac{1-q}{t}\right)f_i} = \frac{f_i}{1-af_i}
\end{equation}
where, for convenience, we use the earlier definition of $a$ and
will often denote this as simply $\tilde{f}_i$. Using the same
approach as in equations (\ref{eqn:derivative-1-pi}) through
(\ref{eqn:deriv-p-end}) we obtain
\begin{displaymath}
\frac{\partial \tilde{p}_i(t)}{\partial t}
=\frac{\tilde{p}_i(t)}{t^2}\left[\tilde{f}_i(t)- \langle
\tilde{f}\rangle(t)\right]
\end{displaymath}
which again, using the same arguments on the aggregation of
states, leads to
\begin{displaymath}
\frac{\partial \tilde{p}_A(t)}{\partial t} =
\frac{\tilde{p}_A(t)}{t^2}\left[\tilde{f}_A(t) -
\langle\tilde{f}\rangle(t)\right].
\end{displaymath}
Note that the only difference between this and the earlier result
is that every occurrence of $p_i$ and $\hat{f}_i$ is replaced with
a $\tilde{p}_i$ and $\tilde{f}_i$, respectively.

\subsubsection{Second Moments}
Proceeding in the same fashion as in (\ref{eqn:deriv-<fhat>Omega})
through (\ref{eqn:derivfhatAend}) we obtain the result
\begin{eqnarray}
\frac{\partial {\langle \tilde{f} \rangle}_\Omega}{\partial t} &=&
\frac{{\langle
\tilde{f}^2\rangle}_\Omega-{\langle\tilde{f}\rangle}_\Omega^2+(1-q){\langle
\tilde{f}^2\rangle}_\Omega}{t^2}\nonumber\\
&=&\frac{\tilde{\sigma}_\Omega^2}{t^2} +
\frac{(1-q){\langle\tilde{f}^2\rangle}_\Omega}{t^2}.\label{eqn:derivftildeOmega}
\end{eqnarray}
Scale invariance in second moments with the Tsallis constraint
Type 1 is indicated by
\begin{eqnarray}
\frac{\partial \tilde{f}_A}{\partial t}&=& \frac{{\langle
\tilde{f}^2\rangle}_A - {\langle\tilde{f}\rangle}_A^2
+(1-q){\langle
\tilde{f}^2\rangle}_A}{t^2}\nonumber\\
&=&\frac{\tilde{\sigma}^2_A}{t^2}+ \frac{(1-q){\langle
\tilde{f}^2\rangle}_A}{t^2}\label{eqn:derivftildeAend}
\end{eqnarray}
where (\ref{eqn:derivftildeOmega}) and (\ref{eqn:derivftildeAend})
are similar to (\ref{eqn:deriv-fhat-Omega}) and
(\ref{eqn:derivfhatAend}) except that, as before, every occurrence
of $q-1$ and $\hat{f}$ has been replaced with a $1-q$ and
$\tilde{f}$, respectively.

\subsection{Probabilities and Energy Relations}
A number of additional forms of symmetry become evident when we
examine the various relationships between probabilities and
transformed energy values. First, suppose we wish to define
$p_i(t)$ in terms of $\hat{f}_i$. Using the energy transformation
function in (\ref{eqn:def-fhat}) it is easy to see that
\begin{equation}\label{eqn:ffromfhat}
f_i = \frac{\hat{f}_i}{1 + \left(\frac{1-q}{t}\right)\hat{f}_i} =
\frac{\hat{f}_i}{1 - a\hat{f}_i}
\end{equation}
where again we use $a = \left(\frac{q-1}{t}\right)$ for notational
convenience. Note that the $q-1$ in (\ref{eqn:def-fhat}) has been
replaced with $1-q$ in (\ref{eqn:ffromfhat}).

Note that (\ref{eqn:def-ftilde}) has a similar form as
(\ref{eqn:ffromfhat}) where $f_i$ is defined in terms of
$\hat{f}_i$, and hence it follows that
\begin{displaymath}
f_i = \frac{\tilde{f}_i}{1 + \left(\frac{q-1}{t}
\right)\tilde{f}_i} = \frac{\tilde{f}_i}{1 + a\tilde{f}_i}
\end{displaymath}
and has the same form as the definition of $\hat{f}_i$ in
(\ref{eqn:def-fhat}).

Now, substituting (\ref{eqn:ffromfhat}) into the form for $p_i(t)$
in (\ref{eqn:defTallis-pi}), we get
\begin{eqnarray*}
p_i(t) &=& \frac{\left[1 + a\left(\frac{\hat{f}_i}{1-a\hat{f}_i}\right) \right]^{\frac{1}{1-q}}}{Z_q}\nonumber\\
&=&\frac{\left[\frac{1}{1-a\hat{f}_i}\right]^{\frac{1}{1-q}}}{Z_q}\\
&=&\frac{\left[1 +
\left(\frac{1-q}{t}\right)\hat{f}_i\right]^{\frac{1}{q-1}}}{Z_q}
\end{eqnarray*}
which, interestingly, is exactly the same form as in
(\ref{eqn:def-ptilde}) above where the Type 1 constraint was used.
This also clearly implies that
\begin{eqnarray*}
\lefteqn{Z_q = \sum_{i=1}^W\left[1 + \left(\frac{q-1}{t}
\right)f_i \right]^{\frac{1}{1-q}}}\hspace{40pt}\nonumber\\
&=& \sum_{i=1}^W\left[1 + \left(\frac{1-q}{t} \right)\hat{f}_i
\right]^{\frac{1}{q-1}}
\end{eqnarray*}
which further suggests another important relationship described
below. Before this relationship is shown, however, a little more
background into the use of other constraints is in order.

The expression $p^q_i$ arise in a number of instances, in
particular, in the definition of $S_q$ itself and so has a very
fundamental character.  It also arises in other fundamental
relationships concerning conditional probabilities and Shannon
Additivity (see \cite{Curado91}).  It also became evident that it
was useful to incorporate them in the constraints that define the
stationary probability.  The ``first choice'' using Tsallis'
notation, $\sum_i\tilde{p}_if_i = U^{(1)}$ lead to problems.  The
``second choice'', \ie $\sum_ip^q_if_i = U^{(2)}$ resolved some of
them. Tsallis summarizes:
\begin{quote}
The first choice was very little used in the literature because
quite quickly it became obvious that it could not solve relevant
mathematical difficulties existing in the approach of anomalous
phenomena such as L\'{e}vy superdiffusion. The second choice has
been profusely used in the literature, and was not quickly
abandoned because the deep physical reason for the generalization
$(q \neq 1)$ was not transparent. But, in the light of recent
developments [ ] showing the relationship of the formalism with a
possible violation of the usual ergodic mixing hypothesis,
features like the $q$-expectation value of unity not being equal
to one became clearly unacceptable. Then naturally emerged the
third choice, which we believe to be fully satisfactory\ldots
\end{quote}
\cite{Tsallis98}.  We leave the implications of the third choice
to future efforts.

Tsallis notes the effects of the exponent $q$ in the type 2
constraint as they ``[privilegiate] the {\em rare} and the {\em
frequent\/} events'' depending on whether $q<1$ or $q>1$,
respectively \cite[p.535]{Tsallis98}.  But this notion of shifting
the probability weight of different energy values is, in some
sense, equivalent to transforming the energy values themselves as
the following lemma illustrates.
\begin{lemma}\label{lem:pqf--barfhat}
For all $q > 1$,
\begin{displaymath}
\sum_ip^q_if_i = Z_q^{1-q}\sum_ip_i\hat{f}_i = Z_q^{1-q}\langle
\hat{f} \rangle.
\end{displaymath}
\end{lemma}
\begin{proof}
It follows from the definition of $p_i$, that for all $i$,
\begin{equation}\label{eqn:pqlemmaproof}
p^q_if_i = \left(\frac{\left[1 + \left(\frac{q-1}{t} \right)
f_i\right]^{\frac{1}{1-q}}}{Z_q}\right)^qf_i.
\end{equation}
Now observe that $\frac{q}{1-q} = \frac{1}{1-q} - 1$. Consequently
for all $i$,
\begin{eqnarray*}
p_i^q &=& \frac{\left[1 + \left(\frac{q-1}{t} \right)
f_i\right]^{\frac{1}{1-q}}}{Z_qZ_q^{q-1}\left[1 +
\left(\frac{q-1}{t} \right) f_i\right]}\\
&=& \frac{p_i}{Z_q^{q-1}\left[1 + \left(\frac{q-1}{t} \right)
f_i\right]}.
\end{eqnarray*}
Substituting this into (\ref{eqn:pqlemmaproof}) and simplifying we
get
\begin{displaymath}
p_i^qf_i = Z_q^{1-q}\,\,p_i\hat{f}_i
\end{displaymath}
and by summing over all $i$, the result follows.
\end{proof}

\noindent It is worth noting that this is consistent with the
Legendre structure as noted in \cite[p.539]{Tsallis98} where the
temperature $t \equiv 1/(k\beta)$.

Finally, it is worthwhile to investigate the relationship between
the constraint Type 1 and the energy transformation function.
Using a similar approach as before, we state the following lemma:
\begin{lemma}\label{lem:Zprimetildef}
For all $q<1$,
\begin{displaymath}
Z^{\prime 1-q}\sum_{i=1}^W\tilde{p}_if_i =
\sum_{i=1}^W\tilde{p}_i^q\tilde{f}_i
\end{displaymath}
\end{lemma}
\begin{proof}
This follows using the same approach as in Lemma
\ref{lem:pqf--barfhat}.
\end{proof}

\section{The Transformed Energy Landscape}
\label{sec:TransformedEnergy}

\begin{figure}[htb]
\vspace{.75in}
\centerline{\includegraphics[width=3.5in]{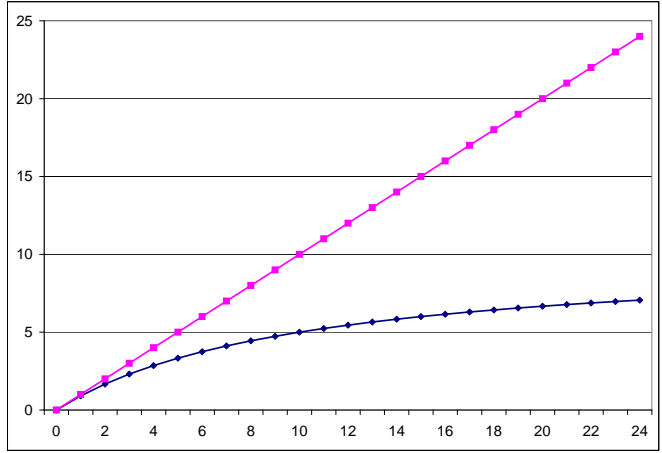}}
\vspace{-.8in} \caption{Plot of $\hat{f}$ versus $f$.  The
transformation of the energy landscape is based on $q=1$ (with
$\blacksquare$ on straight line) and $q=2$ (with $\blacklozenge$
below the straight line) for a fixed value of
$t=10$.}\label{fig:energytrans}
\end{figure}

\begin{figure}[htb]
\vspace{.75in}
\centerline{\includegraphics[width=3.5in]{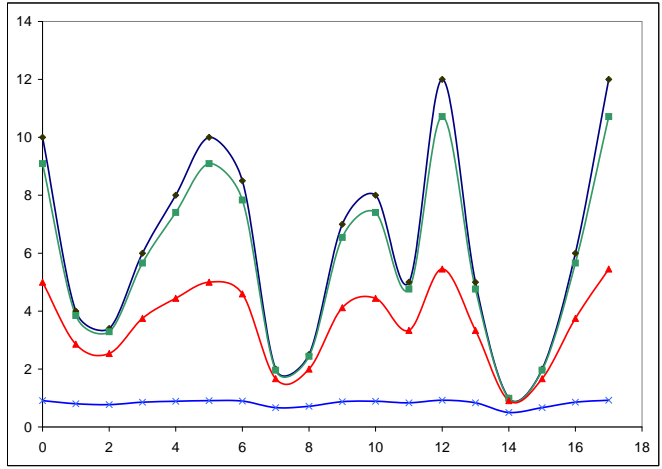}}
\vspace{-.65in} \caption{The energy landscapes of $f$ (with
$\blacklozenge$ at top), and for $q=2$, the values of $\hat{f}$ at
$t=100$ (with $\blacksquare$, 2nd curve), at $t=10$ (with
$\blacktriangle$, 3rd curve) and at $t=1$ (with $\times$ at
bottom).}\label{fig:landscapes}
\end{figure}

The appearance of the form for $\hat{f}_i$ in (\ref{eqn:def-fhat})
raises a number of intriguing issues and possibilities.  In
exploring these issues, it is helpful to gain some sense of what
happens to the values of $\hat{f}_i$ relative to the values of
$f_i$ for different values of the entropic parameter $q$ and
temperature $t$. First note that for $q=1, \ \hat{f}_i = f_i$ for
all $i$ and $t>0$ and produces the straight line depicted in
Figure~\ref{fig:energytrans}. For values of $q>1$, values of
$\hat{f}_i$ are bounded above in that $\lim_{f\rightarrow
\infty}\hat{f} = \frac{t}{q-1}$ and produces a monotonically
increasing curve depicted in Figure~\ref{fig:energytrans} (see
also Lemma \ref{lem:rankorder}). The $q$-transformation reduces
each value of energy in rough proportion to its magnitude. This
has the effect of `flattening out' the energy landscape when $q>1$
as depicted in Figure~\ref{fig:landscapes} for several different
temperatures with $q=2$.
It is worthwhile to note that for values of $q\geq 1$, the rank
order of all values of $f$ and $\hat{f}$ are preserved as the
following lemma shows.

\begin{lemma}\label{lem:rankorder}
For all values of $a>0$ (\ie $q>1$ and $t
> 0$) and for all positive reals $x$ and $y$,
\begin{displaymath}
x > y \Leftrightarrow \frac{x}{1+ax} > \frac{y}{1+ay}.
\end{displaymath}
\end{lemma}

\begin{proof}
For positive $a,x$ and $y,\ x>y$ if and only if $x+axy > y + axy$.
Simplifying and dividing each side by $1+ay$ and $1+ax$ (on the
right), the result follows.
\end{proof}

\vspace{6pt} \noindent It is also easy to see from
Figure~\ref{fig:landscapes} and Lemma~\ref{lem:rankorder} that any
curve at a given temperature majorizes the corresponding curve at
a lower temperature.

The net effect of the energy transformation as indicated in the
preceding can be summarized in the following way: it smoothes out
the landscape, \ie reduces the relief of the landscape and lowers
the energy values yet preserves the essential relational features
of the landscape (in terms of rank order).

\subsection{Relating Exponentials and Powerlaws}
This energy landscape transformation provides several alternative
perspectives on power-law distributions.  The flattened landscapes
in Figure \ref{fig:landscapes} may explain the heavy tail
distributions associated with complex phenomena---the flatter
energy landscape makes it easier, in some sense, to move to higher
values of $\hat{f}_i$ in the energy spectrum.  For example, in
using the classical Metropolis Algorithm and its associated
exponential form using values of $\hat{f}_i$, uphill moves are
more probable and this leads to a heavy-tailed ``steady-state''
distribution.

To make this precise, the following theorem shows how this energy
transformation in effect parameterizes an exponential distribution
and permits it to change into a power-law distribution without the
necessity of taking limits.  This is the opposite of what Tsallis
describes as his ``$q$-exponential'' (see \eg \cite[p.
537]{Tsallis98}) where he states a power-law distribution that in
the limit becomes exponential (recall that $\lim_{q\rightarrow
1}p_i = \pi_i$). Thus, in the limit a {\em power-law} takes on an
{\em exponential} form.  Here, we use the standard definition of
an exponential form which asymptotically becomes a power-law.
\begin{theorem}\label{thm:asymptotics}
Let $a>0$ and $x$ be such that $ax^\gamma > 1$ and define
\[\hat{x}=\frac{x^\gamma}{1+ax^\gamma}\]
for some $\gamma > 0$, then for any $\lambda > 0$
\begin{equation}\label{eqn:limitthm}
e^{-\lambda\hat{x}}-C_1\ \sim \ C_2x^{-\gamma}
\end{equation}
as $ax\rightarrow\infty$ where the constants $C_1= e^{-\lambda/a}$
and $C_2 = a^{-2}\lambda\, e^{-\lambda/a}$.
\end{theorem}
\begin{proof}
First note that
$\lim_{x\rightarrow\infty}\lambda\hat{x}
= \lambda/a$, 
hence, for any fixed $a>0$ both sides of (\ref{eqn:limitthm}) have
limits of 0 as $x\rightarrow \infty$. To show that
$e^{-\lambda\hat{x}}-e^{-\lambda/a}$ decreases to zero
asymptotically as $1/x^\gamma$ (within a constant), first note the
identity for any real $x$ and $\gamma > 0$,
\begin{equation}\label{eqn:expansion:1+x}
\frac{1}{1+x^\gamma}=
1-x^\gamma+x^{2\gamma}-x^{3\gamma}+x^{4\gamma}-\cdots.
\end{equation}
Dividing the numerator and denominator of $\hat{x}$ by
$ax^\gamma$, we obtain
\begin{equation}\label{eqn:expansion-xhat}
\hat{x}= \frac{x^\gamma}{1+ax^\gamma} =
\frac{1}{a}\left(\frac{1}{1+\frac{1}{ax^\gamma}} \right).
\end{equation}
Using the general relationship in (\ref{eqn:expansion:1+x}) in the
parenthesis in (\ref{eqn:expansion-xhat}) and substituting into
$e^{-\lambda\hat{x}}$ yields
\begin{eqnarray}
e^{-\lambda\hat{x}}
&=&\exp\left\{-\frac{\lambda}{a}\left[1-\frac{1}{ax^\gamma}+\frac{1}{(ax^\gamma)^2}
\cdots\right]\right\}\nonumber\\
&=&\exp\left\{-\frac{\lambda}{a}+\frac{\lambda}{a^2x^\gamma}-\frac{\lambda}{a^3x^{2^\gamma}}
+
\cdots\right\}\nonumber\\
&=&e^{-\lambda/a}\exp\left\{\frac{\lambda}{a^2x^\gamma}-\frac{\lambda}{a^3x^{2^\gamma}}
+ \cdots\right\}.\label{eqn:exponentseries}
\end{eqnarray}
Note that since $\gamma > 0$ and $ax^\gamma>1$, the series in
(\ref{eqn:exponentseries}) is absolutely convergent (using the
ratio test), hence converges. Substituting a Taylor Series
expansion of the exponential in (\ref{eqn:exponentseries}) (the
second factor) yields
\begin{equation}
e^{-\lambda\hat{x}}
=e^{-\lambda/a}\left[1+\left(\frac{\lambda}{a^2x^\gamma}-\frac{\lambda}{a^3x^{2^\gamma}}
+ \cdots \right) +
\frac{1}{2}\left(\frac{\lambda^2}{a^4x^{2\gamma}}+\cdots
\right)+\cdots\right]\label{eqn:exptaylor}
\end{equation}
where we note that the higher order terms of the series in
(\ref{eqn:exptaylor}) all have powers higher than $a^4x^{2\gamma}$
in the denominators (these are not shown).  To show that
$e^{-\lambda\hat{x}}$ falls off according to a power law, we
evaluate the expression
\begin{equation}\label{eqn:lim}
\lim_{ax\rightarrow\infty}
\frac{\left(e^{-\lambda\hat{x}}-e^{-\lambda/a}\right)}{a^{-2}\lambda\,
e^{-\lambda/a}x^{-\gamma}}
\end{equation}
to assess its asymptotic behavior and the rate at which it
approaches its limiting value (if any). Substituting
(\ref{eqn:exptaylor}) for the numerator of (\ref{eqn:lim}) and
multiplying the numerator by $x^\gamma$ (from the $x^{-\gamma}$ in
the denominator) and reordering the terms of the Taylor Series
expansion to indicate only those with powers of $a^4x^{2\gamma}$
or less yields
\begin{displaymath}
x^\gamma
e^{-\lambda/a}\left[1+\frac{\lambda}{a^2x^\gamma}-\frac{\lambda}{a^3x^{2\gamma}}
+ \frac{\lambda^2}{2a^4x^{2\gamma}}\cdots \right]-x^\gamma
e^{-\lambda/a}
\end{displaymath}
\vspace{-12pt}
\begin{displaymath}
=\frac{\lambda\, e^{-\lambda/a}}{a^2}\left[1-\frac{1}{ax^\gamma} +
\frac{\lambda}{2a^2x^\gamma}+\cdots \right]
\end{displaymath}
in the numerator.  Consequently,
\begin{eqnarray*}
\lefteqn{\lim_{ax\rightarrow\infty}
\frac{x^\gamma\left(e^{-\lambda\hat{x}}-e^{-\lambda/a}\right)}{a^{-2}\lambda\, e^{-\lambda/a}}\hspace{20pt}}\nonumber\\
&=& \lim_{ax\rightarrow\infty}\left[1-\frac{1}{ax^\gamma} +
\frac{\lambda}{2a^2x^\gamma}+\cdots \right] \nonumber \\
&=&\lim_{ax\rightarrow\infty}\left[1-E(a,x)\right] = 1
\label{eqn:lim=1}
\end{eqnarray*}
where the discrepancy $E(a,x) = \frac{1}{ax^\gamma} -
\frac{\lambda}{2a^2x^\gamma} \rightarrow 0$ as
$ax\rightarrow\infty$ and hence (\ref{eqn:limitthm}) follows.
\end{proof}

\begin{figure}[htb]
\vspace{.75in}
\centerline{\includegraphics[width=3.5in]{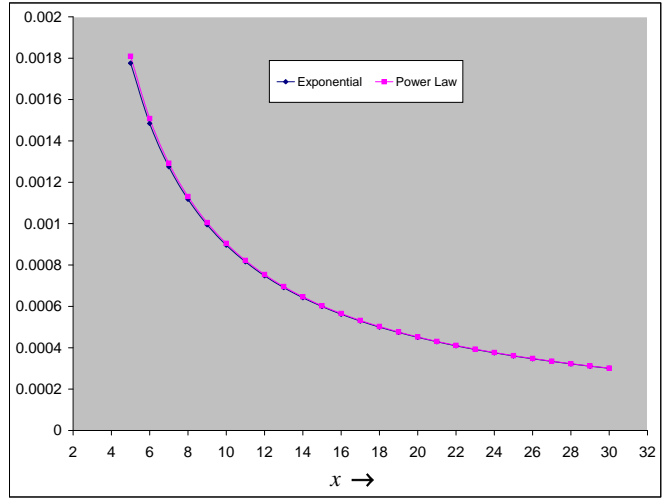}}
\vspace{-.7in} \caption{Plots of
$e^{-\lambda\hat{x}}-e^{-\lambda/a}$ (the ``exponential'' series)
and $\lambda\,e^{-\lambda/a}a^{-2}x^{-\gamma}$ (the ``power law''
series) for $a = 10, \lambda = \gamma =
1$.}\label{fig:exp-vs-powerlaw}
\end{figure}

Figure \ref{fig:exp-vs-powerlaw} compares the decays of
$e^{-\lambda\hat{x}}-e^{-\lambda/a}$ and
$\lambda\,e^{-\lambda/a}/(a^2x^\gamma)$ for the given values of
$a$ and $\lambda$ and shows that the curves are virtually
coincident even for the relatively low values of $x$ indicated in
the figure.

\subsection{A Recursive Transformation}
The form of the $q$--transformation of energy also permits stating
an interesting recursive feature that may be useful in relating
the {\em level\/} of aggregation to the entropic parameter. Recall
that the scale invariant properties indicated in
(\ref{eqn:deriv-pA}) and (\ref{eqn:derivfhatAend}) are based on
aggregating energy levels.  From these aggregations, several
scale-invariant relationships can be defined involving the
transformed energy values $\hat{f}_i$ defined in
(\ref{eqn:def-fhat-A}).  But this value is associated with an {\em
aggregation of particles}, each with its own value of ``energy''
where this aggregation may actually result from a {\em series of
successive} aggregations because the $q$--transformation has this
recursive aspect.  This recursive feature of the energy
transformation in (\ref{eqn:def-fhat}) is based on the following
relationship where we substitute $\hat{f}_i$ for $f_i$ in
(\ref{eqn:def-fhat-A}).  This results in
\begin{equation}\label{eqn:fhat[k+1]}
\hat{f}_{i,k+1}^\gamma =
\frac{\hat{f}_{i,k}^\gamma}{1+a\hat{f}_{i,k}^\gamma}.
\end{equation}
For consistency in notation, let $\hat{f}_{i,0}^\gamma \equiv
f_i^\gamma$ and it follows that $\hat{f}_{i,1}\equiv \hat{f}_i$
for $\gamma = 1$ as defined earlier (see (\ref{eqn:def-fhat})).
Using these relationships, the following lemma holds.
\begin{lemma}\label{lem:fhatinduction}
The $k^{\mbox{\tiny th}}$ energy transformation $\hat{f}_{i,k}$
raised to any power $\gamma > 0$ is related to $f_i$ as follows:
\begin{equation}\label{eqn:recursiveform}
\hat{f}_{i,k}^\gamma = \frac{f_i^\gamma}{1+ k\,af_i^\gamma}
\end{equation}
\end{lemma}
\begin{proof}
This is an obvious implication of (\ref{eqn:fhat[k+1]}) and the
use of the induction method.
\end{proof}

This suggests there is some equivalence between the number of
aggregations and the value of the entropic parameter $q$.  That
is, the coefficient of $f_i$, $ka$, in the denominator of
(\ref{eqn:recursiveform}) corresponds to a higher value of the
entropic parameter.  Successive (or larger?) aggregations of
energy are, in some sense, equivalent to larger values of $q$.
 These asymptotic results and the recursion described in Lemma
\ref{lem:fhatinduction} may provide new perspectives on power laws
and the nature of critical phenomena which are discussed below.

\section{Conclusion and Future Research}\label{sec:conclusion}
This article has explored how various statistical components of a
non-extensive system exhibit forms of scale invariance.  By
aggregating energy levels and defining certain statistical
quantities for these aggregated sets in appropriate ways, such as
its stationary probability and expected energy levels, various
operations on these quantities result in identical mathematical
forms as for the corresponding quantities associated with
individual energy states {\em provided they are based on a
transformed energy value}. This energy transformation appears in a
number of scale invariant and symmetry relationships.

The manifestation of the energy transformation in these scale
invariant forms is a surprising element and bears further
investigation.  The manner in which it appears suggests that scale
invariance exists only by virtue of an energy transformation
function.  This suggests a view that information (entropy) loss
may occur between aggregated states where there has {\em already}
been some information loss due to prior aggregations. Thus, if one
can make the leap from aggregating a system of components to
aggregating their associated energy levels, it may provide insight
into a number of critical phenomena relating to power-law
behaviors.  Thus, aggregated system components may further combine
to form yet larger aggregations where there is a further loss of
entropy due to non-extensivity.  But from the recursive property
in \ref{lem:fhatinduction}, there is in some sense an equivalence
between a {\em succession of aggregations} and changes in
temperature and/or the entropic parameter $q$.  Thus, the level of
aggregations in particular parts of a large systems can be seen as
equivalent to differences in the temperature, heat capacity and so
forth between these different parts of a large system in a
far-from-equilibrium regime.

This may also provide some insight into why certain types of
systems have particular values of critical exponents. For example,
in the border between order and disorder, information loss may
occur in connection with how inelastic collisions manifest
themselves during a phase transition.  But successive aggregations
may happen for only a certain fraction of the energy states in a
system or the tendency towards aggregation may change with the
size of aggregated states. Successive aggregations and the
attendant loss of entropy (when $q> 1$) may then lead to a certain
distribution in terms of the size of these aggregated states
indicated by a power-law. The relative frequency of the size of
aggregated states and their dependance on macroscopic properties
could then explain the particular values of critical exponents as
reflecting the average number of aggregations in large systems. In
this sense, the universality of these critical exponents suggests
that there may be a {\em typical\/} number of successive
aggregations with a distribution of this number that reflects and
explains the non-integer values of critical exponents.

A number of these issues are further explored and developed in
\cite{Fleischer-SETNSM} where a generalization of Tsallis' entropy
definition sheds some additional light on these energy
transformations.  Transformation {\em families} are explored as
well as recursive relationships similar to those described above.
In addition, we use the generalization of Tsallis' entropy to
provide more general ways to state power law distributions in
exponential forms, a reverse approach to Tsallis where he provides
a mechanism where a power law form can be made asymptotically
equal to an exponential form.  This also provides clues as to how
to take fuller advantage of these results in modeling complex
systems and developing simulation models.  New approaches for the
simulation of complex systems and ways in which to overcome the
so-called ``broken ergodicity problem'' are also explored (see
\cite{Fleischer-BrokenErgodicity}).

\begin{center}
\section*{\large Acknowledgements}
\end{center}
This research was supported as an Independent Research and
Development project at the Johns Hopkins University Applied
Physics Laboratory.  The author would like to thank Susan Lee,
Donna Gregg and William Blackert for their support and
encouragement during this project and Drs. Constantino Tsallis,
I-Jeng Wang and Alan Weiss for their very useful and helpful
suggestions.
\clearpage
\bibliography{SISRNSM,FleischerGeneral}

\end{document}